\documentclass{Interspeech2024}

\usepackage{adjustbox}
\usepackage{graphicx}
\usepackage{multirow}
\usepackage{cite}
\usepackage{colortbl}
\usepackage{xcolor}
\usepackage{booktabs}
\usepackage{hyperref}
\usepackage{tabularx}
\usepackage{makecell}
\usepackage{pifont}
\usepackage{bbding}
\usepackage{float}
\usepackage{subcaption}
\usepackage{xcolor}
\usepackage{colortbl}
\usepackage{hyperref}

\interspeechcameraready

\title{TS3-Codec: Transformer-Based Simple Streaming Single Codec}

\name[]{Haibin}{Wu}
\name[]{Naoyuki}{Kanda}
\name[]{Sefik}{Emre Eskimez}
\name[]{Jinyu}{Li}

\address{
  Microsoft, USA
  }
\email{haibinwu@microsoft.com}

\keywords{Transformer, neural audio codec, streaming, low bitrate, low token rate, computation efficiency}

\begin{document}

\maketitle

\begin{abstract}

Neural audio codecs (NACs) have garnered significant attention as key technologies for audio compression as well as audio representation for speech language models.
While mainstream NAC models are predominantly convolution-based, the performance of NACs with a purely transformer-based, and convolution-free architecture remains unexplored.
This paper introduces TS3-Codec, a \textbf{\underline{T}}ransformer-Based \textbf{\underline{S}}imple \textbf{\underline{S}}treaming \textbf{\underline{S}}ingle Codec.
TS3-Codec consists of only a stack of transformer layers with a few linear layers, offering greater simplicity and expressiveness by fully eliminating convolution layers that require careful hyperparameter tuning and large computations.
Under the streaming setup, 
the proposed TS3-Codec achieves comparable or superior performance
compared to the codec with state-of-the-art convolution-based architecture
while requiring only 12\% of the computation and 77\% of bitrate. Furthermore, it significantly outperforms 
the convolution-based codec
when using similar computational resources.\footnote{
If authors wish to compare their model with TS3-Codec, please contact the first author via email to request reconstructed audio samples.}

\end{abstract} 

\section{Introduction}
\label{intro}

Neural audio codec (NAC) 
is a technique to compress audio signals into a sequence of discretized codes for efficient data storage and transmission 
\cite{kim2024neural, wu-etal-2024-codec, shi2024espnet, wu2024codec, mousavi2024dasb}.
More recently, NAC has also gained significant attention 
as a key technology for speech language modeling (SLM) \cite{audiolm,wang2023neural,wang2024speechx}.
By converting continuous audio into discrete codes, large language modeling (LLM) techniques—already highly successful in text processing—is able to be applied to versatile speech processing \cite{wu2024towards}.

Numerous high-performance NACs have been proposed\footnote{\url{https://github.com/ga642381/speech-trident}}, addressing various aspects, e.g. better audio quality, bitrate efficiency, and low computational cost.
Most models rely on convolutional layers as the dominant architecture, with only a few \cite{ji2024wavtokenizer, defossez2024moshi} incorporating transformers (or self-attention mechanism) \cite{vaswani2017attention} as intermediate layers within the convolutional encoder-decoder framework. 
However, the performance of a purely transformer-based and convolution-free architecture in NACs remains unexplored.
This study aims to fill the existing gap by developing a NAC exclusively based on transformer models. It leverages the benefits of transformers, such as simplicity in model design and enhanced computational efficiency when compared to convolution-based models.

When the NAC is used as the token representation for SLMs, the following properties are particularly important.
\begin{itemize}
    \item Streaming: Full-duplex communication, where users and machines can speak and respond simultaneously, is a popular and ongoing challenge in the SLM field \cite{defossez2024moshi,dGSLM}. To enable seamless real-time interactions, the codec must support streaming processing, allowing it to encode user speech and generate speech response with low latency.
    \item Single codebook: A single codebook-based model is preferable to a multiple-codebook-based model, such as residual vector quantization (RVQ)~\cite{zeghidour2021soundstream, defossez2022high}, because the latter introduces additional complexity to the architecture of SLMs, such as the combination of auto-regressive and non-autoregressive models \cite{wang2023neural}, the temporal and depth transformers \cite{yu2023megabyte,yang2023uniaudio}, etc.
    \item Low computation: Low-computation NACs enable faster encoding and decoding, reducing computational demands and leaving more computation resources available for SLMs.
    \item Low token rate:  
    Long sequences generally make LLM training slow and unstable. Therefore, it is preferable to use low-token-rate NAC models for SLM.
\end{itemize}

This paper introduces
TS3-Codec (Transformer-Based Simple Streaming Single Codec),
the first attempt to develop a convolution-free, transformer-only NAC.
TS3-Codec consists of only a stack of transformer layers with a few linear layers,
offering greater simplicity and expressiveness by fully eliminating
convolution layers that require careful hyperparameter tuning and large computations. 
The proposed TS3-Codec offers several advantages, namely, streaming capability, low computational requirements, low bitrate, and a single codebook design
while maintaining high audio quality.
In the streaming setup, 
the proposed TS3-Codec delivers comparable or superior performance than convolution-based codecs with just 12\% of the computation and 77\% of bitrate.
TS3-Codec also achieves
significantly better audio quality when using the same computational resources.

\section{Related work}
\label{related}

Neural audio codec models typically consist of an encoder, a vector quantization (VQ) module, and a decoder.
The encoder downsamples the time-domain audio to extract frame-wise audio features, typically with a frame rate of 12.5–100 Hz. 
The vector quantization module — whether single-quantizer vector quantization \cite{van2017neural}, residual vector quantization (RVQ) \cite{zeghidour2021soundstream, defossez2022high}, or scalar quantization (SQ) \cite{mentzer2023finite} — converts each frame-wise audio feature into discrete tokens. 
These discrete tokens can be used for efficient transmission or as input for SLMs.
Finally, the decoder reconstructs the time-domain audio signal from the discrete tokens.
To meet the requirements of SLMs for real-time conversational agents with full-duplex mode, a suitable codec is better to support streaming, low computational complexity, a single codebook, and a low token rate.

\subsection{Streaming}
Encodec \cite{defossez2022high} is one of the pioneers in achieving real-time, high-quality audio coding by utilizing a streaming encoder-decoder architecture with residual vector quantization.
To enable full-duplex operation in SLMs, Mimi \cite{defossez2024moshi} is proposed as a streaming RVQ codec with a 12.5 Hz token rate. 
It employs techniques, e.g. semantic distillation, discriminator-only training without reconstruction loss, and the integration of transformer layers between the encoder and decoder backbone to enhance feature modeling. 
Mimi achieves a low bitrate of 1.1 kbps while maintaining very high speech quality.

\subsection{Single codebook}
A single codebook-based model is preferable to a multiple-codebook-based model, because it offer the simplicity design of the SLMs.
Multiple-codebook-based codecs (e.g. RVQ-based codec) compress speech into multiple streams of tokens. Unlike text tokens, which form a single stream, modeling multiple audio token streams requires a more complex model design for decoding. Though various decoding strategies have been proposed, including decoding different streams in separate steps \cite{audiolm,wang2023neural}, introducing delay patterns across various stream tokens \cite{copet2024simple}, and combining temporal and depth transformers for generation \cite{yu2023megabyte,yang2023uniaudio}, these multi-stream approaches obviously increase the complexity of model design. In contrast, single-stream codecs offer simple model design.

TiCodec \cite{ren2024fewer} and SingleCodec \cite{li2024single} are designed to encode speech using fewer tokens by disentangling time-invariant global information (e.g., speaker timbre and acoustic environment) into a single utterance-level vector, while representing time-varying information (e.g., phonetic content) with a single stream of frame-level tokens. 
The time-invariant utterance embeddings and frame-level token sequences are then combined to reconstruct the audio. 
However, a limitation of these codecs is their reliance on global utterance-level features for decoding, which restricts their streaming capability.
WavTokenizer \cite{ji2024wavtokenizer} employs several techniques, including codebook size optimization, k-means initialization for codebook embeddings, the Vocos decoder \cite{siuzdak2023vocos}, and training on 80,000 hours of data, to achieve high fidelity in a single-codebook low-bitrate codec.
BigCodec \cite{xin2024bigcodec} achieves exceptionally high reconstruction quality at a low bitrate of 1.04 kbps by scaling the convolutional model size to 160M parameters, making it a strong convolutional baseline for our work.


\subsection{Uniqueness and contributions of this work}
This paper presents the first transformer-only codec architecture designed for streaming, featuring low computational complexity and a simple model design.
Previous works have utilized transformers \cite{defossez2024moshi,ji2024wavtokenizer} solely as intermediate layers within predominantly convolutional backbones for feature engineering. 
Furthermore, previous streaming codecs \cite{defossez2024moshi,defossez2022high,wu2023audiodec} rely on residual vector quantization (RVQ), and single-codebook codecs \cite{ren2024fewer,li2024single,ji2024wavtokenizer,xin2024bigcodec} have lacked a streaming design. 
Our work is the first to introduce a single-codebook codec specifically designed for streaming.
This work is also the first to explore single codec designs with 65k and 130k codebook sizes.

\section{TS3-Codec}
\label{method}

\subsection{Model architecture}

\begin{figure*}[h]
    \centering
    \includegraphics[width=2.0\columnwidth]{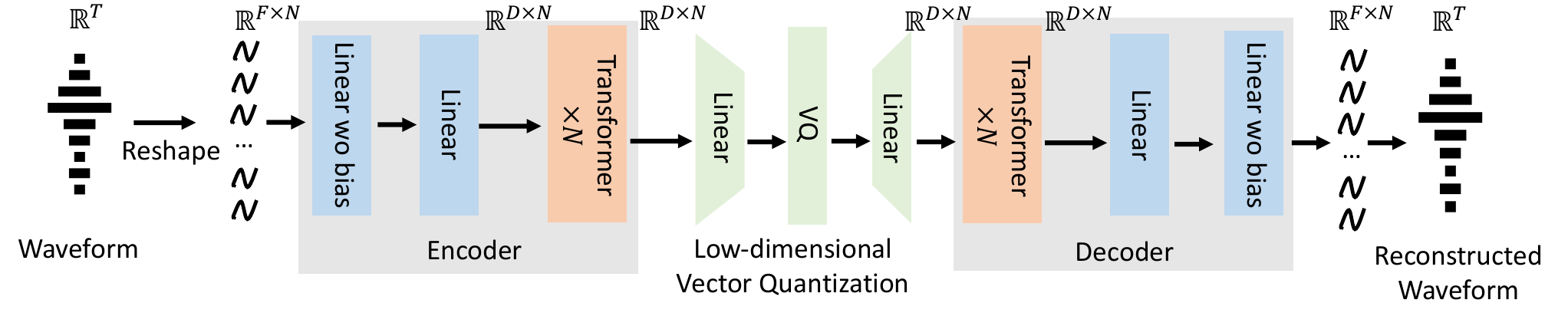}
    \caption{The framework of the proposed TS3-Codec. The transformer layer employs a sliding window attention on the left context, enabling the model to function in a streaming manner with linear (rather than quadratic) complexity relative to the waveform length.}
    \label{fig:framework}
\end{figure*}

TS3-Codec follows the conventional NAC structure, which includes an encoder, 
a quantizer, 
and a decoder, as shown in Figure~\ref{fig:framework}. The model is trained using the generative adversarial network (GAN) framework \cite{goodfellow2020generative}.

The encoder consists of two linear layers and a stack of transformer layers with sliding window attention only on the left context.
The input audio signal with a shape of $\mathbb{R}^T$, where $T$ is the signal length, is first reshaped into a two-dimensional tensor with a shape of $\mathbb{R}^{F\times N}$, 
where $F$ is a window-size and $N=\frac{T}{F}$ is a number of windowed frames.\footnote{For simplicity, we assume $T$ is divisible by $F$. We can pad the original signal to satisfy this assumption.}
Two linear layers without activations (the one close to the waveform is without bias, and the second linear is with bias) are then applied to convert the reshaped windowed frames into a shape of $\mathbb{R}^{D\times N}$.
There is a connector linear layer to connect the two linear layers and the Transformer layers, if needed to fix the dimension mismatch.
Then, a stack of Transformer layers with an embedding size of $D$ is applied
to output an encoder embedding with a shape of $\mathbb{R}^{D\times N}$.
The Transformer layer employs
a sliding window with the size of either 16 or 32 on the left-context for the self-attention operation, meaning only a fixed number of previous frames are considered. 
This design not only ensures that the transformer's computational complexity does not brow up to the quadratic with respect to $N$, but 
also improves the generalization ability of the model for an audio longer than the training data.
If not specified, the encoder has 8 transformer layers, 16 attention heads, and a feed-forward dimension of 4096.

After the encoder module, a factorized VQ layer \cite{van2017neural} discretizes the encoder output in a low-dimensional space. We used a codebook with a codebook dimension of 8 or 16, and a codebook size of 65,536 or 131,072.

The decoder has a symmetric structure with the encoder, consisting of 
a stack of transformer layers with two linear layers.
Unless otherwise specified, the decoder uses the same parameters as the encoder,
namely, 8 transformer layers, 16 attention heads, and a feed-forward dimension of 4096.
The output from the transformer layer has a shape of $\mathbb{R}^{D\times N}$.
It is then converted
into a tensor with a shape of $\mathbb{R}^{F\times N}$ based on two linear layers without activations (the first one is with bias, and the second one doesn't have bias).
Finally, the tensor is reshaped into a single-dimensional tensor
with a shape of $\mathbb{R}^{T}$.\footnote{ Our decoder is largely inspired by the WaveNeXt vocoder \cite{okamoto2023wavenext}, which demonstrated that up-sampling at the final layer using a simple linear layer is sufficient to reconstruct high-quality speech, as opposed to progressive up-sampling through stacked convolution layers. We refer to the implementation in \url{https://github.com/wetdog/wavenext_pytorch}}

\subsection{Design principle}
We chose transformers as the backbone for several potential advantages over convolution-based architectures.
\begin{itemize}
    \item Convolutional layers are well-known for their parameter efficiency and reusability. On the other hand, for models of similar parameter sizes, convolutions typically require significantly more computation than transformers \cite{yuzhang2024lti}.
    Surprisingly, we discovered that the state-of-the-art neural audio codec model, BigCodec \cite{xin2024bigcodec}, with 160M parameters, has a computational cost comparable to that of a \textbf{1-billion-parameter} transformer.\footnote{ At a bitrate of 1k, TS3-Codec (1.6B parameters) requires 60.52G MACs, while BigCodec (160M parameters) requires 61.1G MACs. At a bitrate of 0.6k, TS3-Codec (1.2B parameters) uses 36.34G MACs, while BigCodec (160M parameters) uses 39.6G MACs.}
    The computational cost increases significantly as convolutional neural audio codec models are scaled up, making it impractical to scale these models further.
    \item Convolutions have inherent biases. Convolutions apply fixed weighted-sum weights across all intermediate feature maps across different time stamps, whereas transformers use self-attention, dynamically determining weights tailored to each feature map. Transformers also incorporate positional embeddings, enabling distinct embeddings to be added to different feature maps at different time stamps.
    \item  Transformers offer simplicity in model design. Unlike convolutions, which require careful selection of kernels and up- and down-sampling mechanisms due to their inherent biases (they are sensitive to hyperparameter settings), transformers could avoid these complexities.
\end{itemize}

\subsection{Training objectives}

Similar to most NACs, TS3-Codec is trained based on the GAN framework.
We use multiple losses following BigCodec \cite{xin2024bigcodec}.
\begin{itemize}
    \item Reconstruction Loss: 
     We adopt a multi-scale mel-spectrogram reconstruction loss, calculated using the $L_{1}$ distance in the spectral domain across multiple scales. The mel-spectrogram is closely related to perceptual audio quality.
    \item Least-square GAN loss \cite{mao2017least}:
    Following the BigCodec \cite{xin2024bigcodec}, we employ two types of discriminators to train our model. 
    The first is the Multi-Period Discriminator (MPD), adapted from HiFi-GAN \cite{kong2020hifi}, which captures various periodic patterns in speech signals. The second is the Multi-Scale Short-Time Fourier Transform (MS-STFT) Discriminator, which is used in EnCodec \cite{defossez2022high}. We use the same discriminator configurations as BigCodec. 
    \item Feature loss: we use the $L_{1}$ feature matching loss for the discriminator features.
    \item VQ loss: The codebook is trained using the $L_{1}$ loss, which is calculated between the features before and after quantization, employing a stop-gradient operation \cite{van2017neural}. Following the approach used in BigCodec, we do not use a moving average to update the codebook. To prevent the encoder output from becoming excessively large, a commitment loss with a loss weight as 0.25 is introduced.
\end{itemize}

The loss weights for the reconstruction loss, GAN loss, feature loss are set to 15.0, 1.0, and 1.0, respectively.
Unlike the official BigCodec implementation, we set the VQ loss proportional to the codebook size: 4.0 for a codebook size of 8192, 32.0 for a codebook size of 65,536, and 64.0 for a codebook size of 131,072.
This configuration yielded the best results in our preliminary experiments. Note that the same loss settings are applied during the training of the streaming version of the BigCodec models.




\section{Experimental setup}
\label{setup}

\subsection{Data}
We used Libri-light \cite{kahn2020libri}, which contains 60K hours of speech, to train the codec models.
For evaluation, we used the `test-clean' subset of Librispeech \cite{panayotov2015librispeech}, which consists of 2620 utterances from 40 speakers.

\begin{figure*}[h]
    \centering
    \includegraphics[width=2.0\columnwidth]{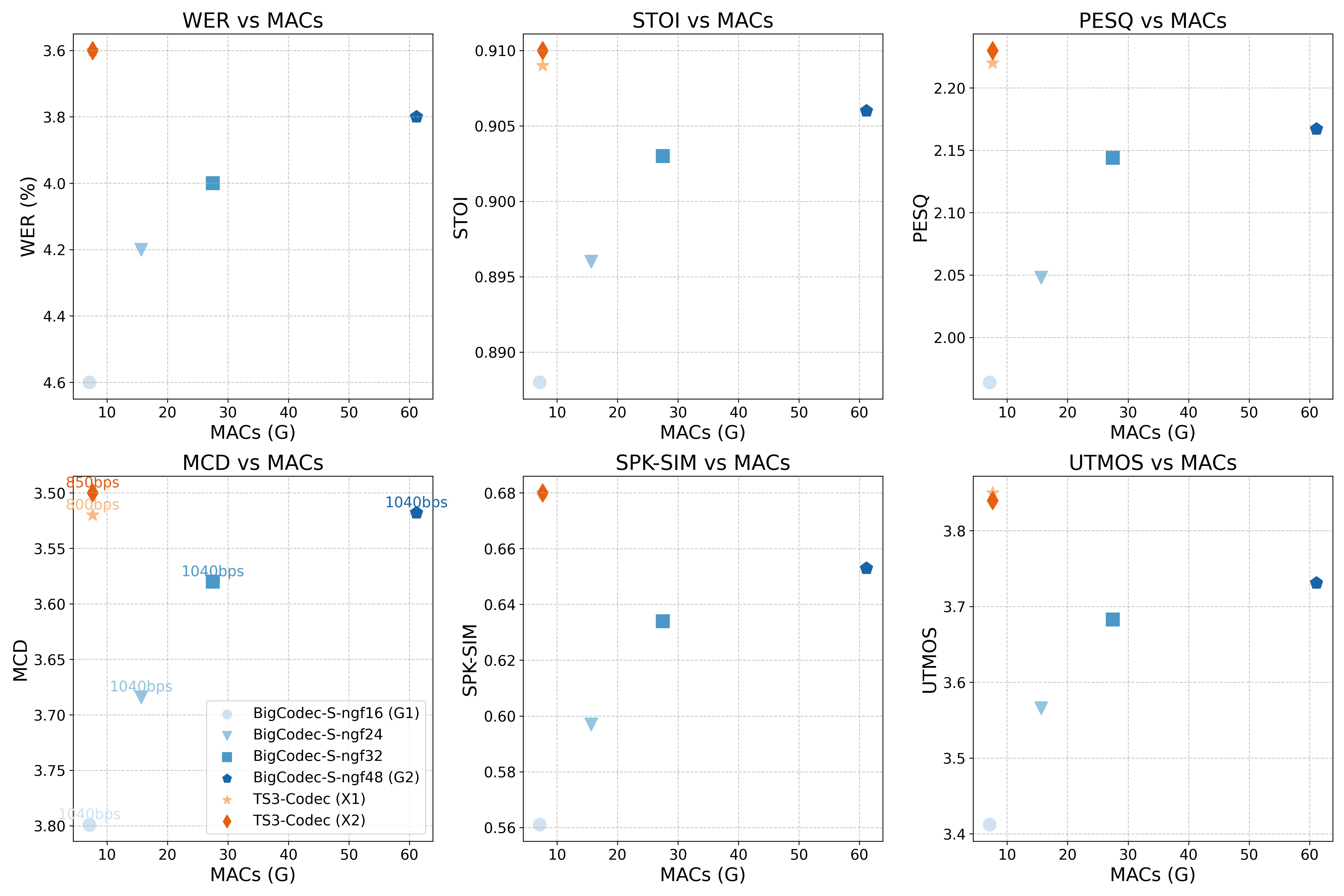}
    \caption{Comparison between BigCodec-S and TS3-Codec (Bitrate $\approx$ 1000 bps). To enhance visualization, the y-axes for WER and MCD are inverted, so that model points in the upper-left corner exhibit the best performance with the least computational cost. ngf is a factor related to the model size of BigCodec-S. }
    \label{fig:1000bps}
\end{figure*}

\begin{figure*}[h]
    \centering
    \includegraphics[width=2.0\columnwidth]{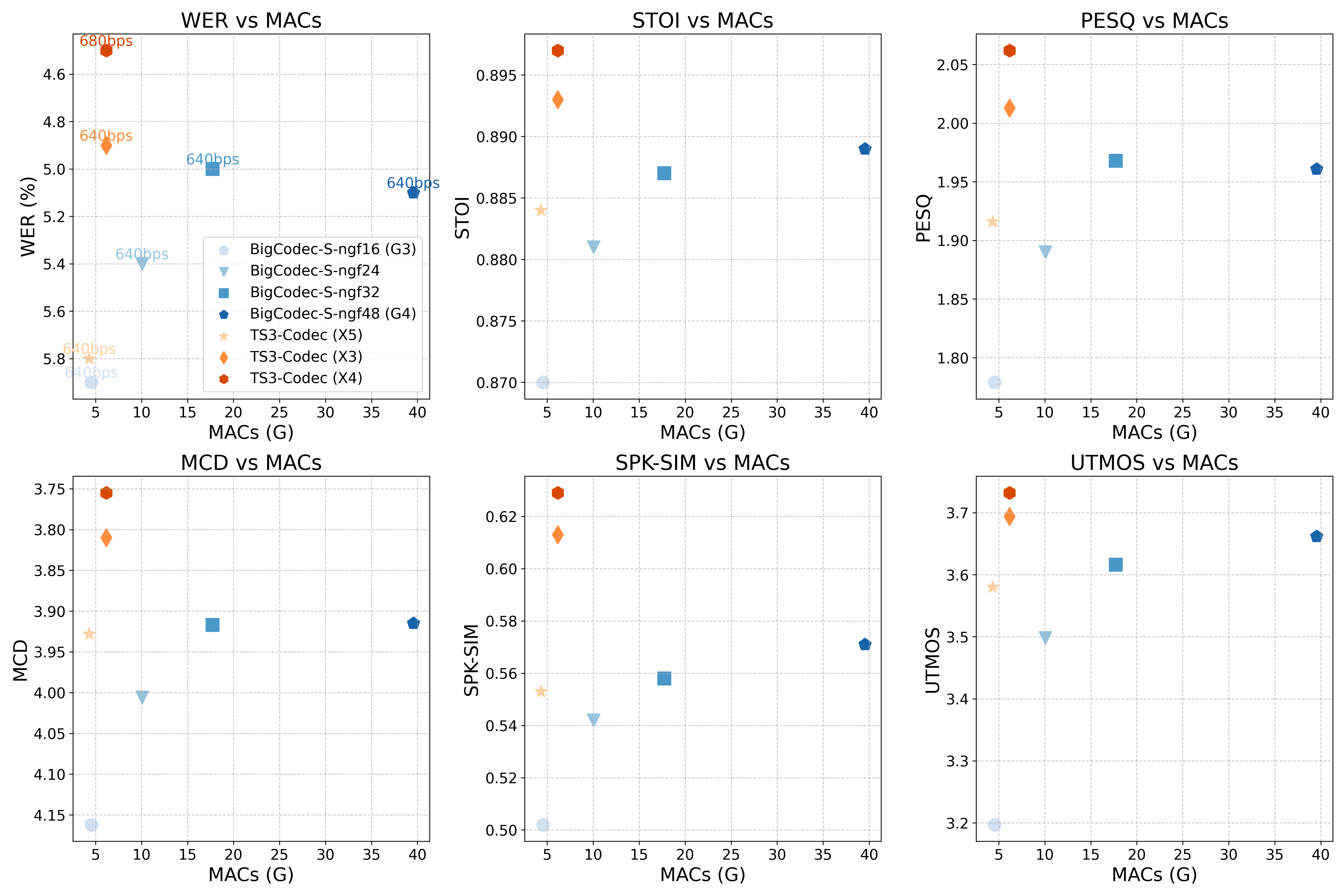}
    \caption{Comparison between BigCodec-S and TS3-Codec (Bitrate $\approx$ 600 bps). To enhance visualization, the y-axes for WER and MCD are inverted, so that model points in the upper-left corner exhibit the best performance with the least computational cost.}
    \label{fig:640bps}
\end{figure*}

\subsection{Evaluation Metrics}
\subsubsection{Complexity and the extracted tokens' property}
\begin{itemize}
    \item Paras: The number of parameters of the codec model.
    \item Multiply-Accumulate Operations (MACs): MACs refer to the fundamental computational operations in neural audio codec models. In this paper, we calculate MACs based on a 1-second audio utterance based on PyFlops\footnote{\url{https://github.com/sovrasov/flops-counter.pytorch/tree/master}}. For modules not supported by PyFlops, such as streaming transformers, we calculated them manually.
    \item Bits Per Second (BPS): measures the number of bits transmitted per second. We used this metric to show the trade-off between transmitted audio quality and compression rate.
    \item Token rate: A metric that indicates the number of tokens required for each second of encoded audio. It is an important measure for speech language modeling.
    \item Frame rate: A metric that shows the number of frames needed to encode each second of audio.
\end{itemize}

\subsubsection{Intelligibility}
\begin{itemize}
    \item Word Error Rate (WER): To evaluate the intelligibility of the reconstructed audio, we employed a HuBERT-based \cite{hsu2021hubert} speech recognition model\footnote{\small \url{https://huggingface.co/facebook/hubert-large-ls960-ft}} to calculate the word error rate (WER). All results are reported as WER percentages.
    \item Short-Time Objective Intelligibility (STOI): assess speech intelligibility by comparing short-time temporal envelopes of clean and degraded speech signals, ranging between 0 and 1, where higher values mean better intelligibility.
\end{itemize}

\subsubsection{Distortion}
\begin{itemize}
    \item  Mel Cepstral Distortion (MCD): MCD measures the difference between two mel-frequency cepstral coefficients (MFCCs) sequences. It is commonly employed in speech synthesis and voice conversion tasks to evaluate the quality of generated speech.  
    \item Perceptual Evaluation of Speech Quality (PESQ): PESQ assesses the quality of speech signals by comparing a degraded speech signal with a reference signal, providing a score that predicts human perception of speech quality
    , where higher values indicate better quality. We used the wide-band PESQ.
\end{itemize}

\subsubsection{Speaker similarity}
To evaluate the speaker similarity (SPK SIM), we used the WavLM-based speaker verification model\footnote{\url{https://github.com/microsoft/UniSpeech/tree/main/downstreams/speaker\_verification}} \cite{chen2022wavlm} to calculate the cosine similarity between original and reconstructed utterances. 

\subsubsection{Naturalness}
We used UTMOS\footnote{\url{https://github.com/sarulab-speech/UTMOS22}} \cite{saeki2022utmos} to evaluate naturalness. 
UTMOS is a neural network-based metric that predicts the Mean Opinion Score (MOS) 
for the naturalness of speech
by learning from pairs of human subjective scores and corresponding utterances, with scores ranging from 1 to 5. 
UTMOS is wildly used as a metric that is highly correlated with human preferences on codec reconstructed utterances \cite{shi2024espnet,ji2024wavtokenizer}.

\begin{table}[t!]
\centering
\fontsize{8}{10}\selectfont
\setlength\tabcolsep{2pt}
\caption{Comparison between baseline codecs. 
\textbf{SEM} represents semantic distillation. 
\textbf{RVQ} and \textbf{single} means residual and single vector quantization, respectively.
\textbf{SA} means self-attention.
}
\label{tab:codec_compare}
\begin{tabularx}{0.48\textwidth}{lcccccc}
\toprule
 \textbf{Codec} & \textbf{SEM} & \textbf{Streaming} & \textbf{VQ type} & \textbf{Architecture} \\
\midrule
Encodec \cite{defossez2022high}         & \ding{55} & \ding{51} & RVQ & Conv + LSTM \\
DAC \cite{kumar2024high}             & \ding{55} & \ding{55} & RVQ & Conv \\
SpeechTokenizer \cite{zhang2023speechtokenizer} & \ding{51} & \ding{55} & RVQ & Conv + LSTM \\
Mimi \cite{defossez2024moshi}            & \ding{51} & \ding{51} & RVQ & Conv + Transformer \\
BigCodec \cite{xin2024bigcodec}       & \ding{55} & \ding{55} & Single & Conv + LSTM \\
WavTokenizer \cite{ji2024wavtokenizer}    & \ding{55} & \ding{55} & Single & Conv + LSTM + SA   \\
\bottomrule
\end{tabularx}
\end{table}
\begin{table}[h]
\centering
\fontsize{6}{8}\selectfont
\setlength\tabcolsep{2pt}
\caption{Hyperparameters for different TS3-Codec models. \textbf{T Layer} and \textbf{T Dim} denote the transformer layer number of encodec/decoder, and transformer feed-forward dimensions. \textbf{E-1} and \textbf{E-2} (\textbf{D-1} and \textbf{D-2}) denote the shapes of the two linear layers for the encoder (decoder). \textbf{Window} denotes the sliding attention window size. \textbf{C size} denotes the codebook size}
\begin{tabularx}{0.48\textwidth}{lcccccccc}
\toprule
\textbf{ID} & \textbf{T Layer} & \textbf{T Dim} & \textbf{E-1} & \textbf{E-2} & \textbf{D-1} & \textbf{D-2} & \textbf{Window} & \textbf{C size} \\
\midrule
X1 & 8  & 4096 & 320$\times$768  & 768$\times$1024  & 1024$\times$768  & 768$\times$320  & 32  & 65536 \\
X2 & 8  & 4096 & 320$\times$768  & 768$\times$1024  & 1024$\times$768  & 768$\times$320  & 32 & 131072\\
X3 & 8  & 4096 & 400$\times$1024 & 1024$\times$1024 & 1024$\times$1024 & 1024$\times$400 & 16 & 65536\\
X4 & 8  & 4096 & 400$\times$1024 & 1024$\times$1024 & 1024$\times$1024 & 1024$\times$400 & 16 & 131072\\
X5 & 10 & 2048 & 400$\times$1024 & 1024$\times$1024 & 1024$\times$1024 & 1024$\times$400 & 16 & 65536 \\
\bottomrule
\end{tabularx}
\label{tab:tes_config}
\end{table}

\begin{table*}[h]
\centering
\caption{Comparison between different codec models under around 1000 bps. BigCodec and BigCodec-S denote the non-streaming and streaming versions, respectively.}
\label{tab:1000bps}
\begin{adjustbox}{width=\textwidth}
\begin{tabular}{@{}l|c|cccc|cc|cc|cc|c|c@{}}
\toprule
\textbf{Model Tag} & \textbf{Streaming} & \textbf{Bitrate} & \textbf{\begin{tabular}[c]{@{}c@{}}Codebook \\ Layer\end{tabular}} & \textbf{\begin{tabular}[c]{@{}c@{}}Frame \\ Rate\end{tabular}} & \textbf{\begin{tabular}[c]{@{}c@{}}Token \\ Rate\end{tabular}} & \textbf{MACs} & \textbf{Paras} & \textbf{WER$\downarrow$} & \textbf{STOI$\uparrow$} & \textbf{PESQ$\uparrow$} & \textbf{MCD$\downarrow$} & \textbf{SPK-SIM$\uparrow$} & \textbf{UTMOS$\uparrow$} \\ \midrule
Ground Truth       & -                  & -            & -                                                              & -                                                           & -                                                             & -            & -             & 2.0         & 1.000         & 4.64          & 0.00         & 1.00            & 4.09           \\ \midrule
DAC (A)            & \ding{55}                  & 1500         & 2                                                              & 75                                                          & 150                                                           & 55.6G       & 74.1M        & 7.2         & 0.829         & 1.48          & 4.83         & 0.47            & 1.68           \\
SpeechTokenizer (B1) & \ding{55}                & 1000         & 2                                                              & 50                                                          & 100                                                           & 17.1G       & 103.7M       & 3.9         & 0.768         & 1.21          & 6.30         & 0.33            & 2.32           \\
BigCodec (C)       & \ding{55}                  & 1040         & 1                                                              & 80                                                          & 80                                                            & 67.1G       & 159.4M       & 2.8         & 0.935         & 2.68          & 3.01         & 0.84            & 4.11           \\
WavTokenizer (D1)  & \ding{55}                  & 975          & 1                                                              & 75                                                          & 75                                                            & 6.3G        & 80.6M        & 6.8         & 0.886         & 2.05          & 4.00         & 0.59            & 3.89           \\ \midrule
Encodec (E1)        & \ding{51}                  & 1500         & 2                                                              & 75                                                          & 150                                                           & 5.6G          & 14.9M        & 4.9         & 0.845         & 1.56          & 4.32         & 0.60            & 1.58           \\
Mimi (F1)          & \ding{51}                  & 1100         & 8                                                              & 12.5                                                        & 100                                                           & 8.1G        & 79.3M        & \cellcolor{gray!20}\textbf{3.0} & 0.905         & 2.22 & 3.81         & \cellcolor{gray!20}\textbf{0.73}   & 3.60           \\
BigCodec-S (G1)      & \ding{51}                  & 1040         & 1                                                              & 80                                                          & 80                                                            & 7.1G        & 21.8M        & 4.6         & 0.888         & 1.96          & 3.80         & 0.56            & 3.41           \\
BigCodec-S (G2)      & \ding{51}                  & 1040         & 1                                                              & 80                                                          & 80                                                            & 61.1G       & 159.9M        & 3.8         & 0.906         & 2.17          & 3.52         & 0.65            & 3.73           \\
TS3-Codec (X1)   & \ding{51}                  & 800          & 1                                                              & 50                                                          & 50                                                            & 7.6G         & 203.6M       & 3.6         & 0.909 & 2.22          & 3.52 & 0.68            & \cellcolor{gray!20}\textbf{3.85}  \\
TS3-Codec (X2)   & \ding{51}                  & 850          & 1                                                              & 50                                                          & 50                                                            & 7.6G         & 203.6M       & 3.6 & \cellcolor{gray!20}\textbf{0.910}         & \cellcolor{gray!20}\textbf{2.23}          & \cellcolor{gray!20}\textbf{3.50}         & 0.68            & 3.84           \\ \bottomrule
\end{tabular}
\end{adjustbox}
\end{table*}
\begin{table*}[h]
\centering
\caption{Comparison between different codec models under around 600 bps.}
\label{tab:600bps}
\begin{adjustbox}{width=\textwidth}
\begin{tabular}{@{}l|c|cccc|cc|cc|cc|c|c@{}}
\toprule
\textbf{Model Tag} & \textbf{Streaming} & \textbf{Bitrate} & \textbf{\begin{tabular}[c]{@{}c@{}}Codebook \\ Layer\end{tabular}} & \textbf{\begin{tabular}[c]{@{}c@{}}Frame \\ Rate\end{tabular}} & \textbf{\begin{tabular}[c]{@{}c@{}}Token \\ Rate\end{tabular}} & \textbf{MACs} & \textbf{Paras} & \textbf{WER$\downarrow$} & \textbf{STOI$\uparrow$} & \textbf{PESQ$\uparrow$} & \textbf{MCD$\downarrow$} & \textbf{SPK-SIM$\uparrow$} & \textbf{UTMOS$\uparrow$} \\ \midrule
Ground Truth       & -                  & -            & -                                                              & -                                                           & -                                                             & -            & -             & 2.0         & 1.000         & 4.64          & 0.00        & 1.00            & 4.09           \\ \midrule
SpeechTokenizer (B2) & \ding{55}                & 500         & 1                                                              & 50                                                          & 50                                                           & 17.1G       & 103.7M       & 4.9         & 0.675         & 1.12          & 8.38         & 0.17            & 1.34           \\
WavTokenizer (D2)  & \ding{55}                  & 520          & 1                                                              & 40                                                          & 40                                                            & 3.4G        & 80.9M        & 8.0         & 0.868         & 1.88          & 4.32         & 0.57            & 3.77           \\ \midrule
Encodec (E2)          & \ding{51}                  & 750       & 1                                                              & 75                                                        & 75                                                          & 5.6G        & 14.9M        &  29.0        &   0.770       &  1.23         &  5.66        &   0.25          &   1.25         \\
Mimi (F2)          & \ding{51}                  & 687.5       & 5                                                              & 12.5                                                        & 62.5                                                          & 8.1G        & 79.3M        & \cellcolor{gray!20}\textbf{4.0}         & 0.872         & 1.82          & 4.40         & 0.58            & 3.27           \\
BigCodec-S (G3)    & \ding{51}                  & 640         & 1                                                              & 40                                                          & 40                                                            & 4.6G       & 21.8M       & 5.9         & 0.870         & 1.78          & 4.16         & 0.50            & 3.20           \\
BigCodec-S (G4)    & \ding{51}                  & 640         & 1                                                              & 40                                                          & 40                                                            & 39.6G       & 160.5M      & 5.4         & 0.889         & 1.96          & 3.97         & 0.58            & 3.68           \\
TS3-Codec (X3)   & \ding{51}                  & 640         & 1                                                              & 40                                                          & 40                                                            & 6.2G       & 204.4M      & 4.9         & 0.893         & 2.01          & 3.81         & 0.61            & 3.69           \\
TS3-Codec (X4)   & \ding{51}                  & 680         & 1                                                              & 40                                                          & 40                                                            & 6.2G       & 204.4M      & 4.5         & \cellcolor{gray!20}\textbf{0.897}         & \cellcolor{gray!20}\textbf{2.06}          & \cellcolor{gray!20}\textbf{3.75}         & \cellcolor{gray!20}\textbf{0.63}            & \cellcolor{gray!20}\textbf{3.73}           \\ \bottomrule
\end{tabular}
\end{adjustbox}
\end{table*}

\subsection{Baselines}
\subsubsection{Official checkpoints}
For the baselines, we included official checkpoints from current high-performing codec models as references. 
The comparison of these 6 codec models is shown in Table~\ref{tab:codec_compare}.
\begin{itemize}
    \item Encodec \cite{defossez2022high}: We used the default 24k Hz model\footnote{\url{https://github.com/facebookresearch/encodec}}, with a bitrate setting as 1.5 kbps, which aligns closely with our model's bitrate. During reconstruction, we first upsampled the utterance to 24 kHz, then performed reconstruction using Encodec, and finally downsampled it to obtain the reconstructed utterances at 16 kHz.
    \item DAC \cite{kumar2024high}: We used the 24k Hz model\footnote{\url{https://github.com/descriptinc/descript-audio-codec}}, and adopted similar procedures as for Encodec.
    \item SpeechTokenizer \cite{zhang2023speechtokenizer}: We used the speechtokenizer\_snake model\footnote{\url{https://github.com/ZhangXInFD/SpeechTokenizer}}. We got two bitrate settings, 1k and 0.5k bps, to align with our bitrate settings.
    \item Mimi \cite{defossez2024moshi}: We used the official checkpoint\footnote{\url{https://huggingface.co/kyutai/mimi}} to obtain two bitrate settings: 1.1 kbps and 0.6875 kbps, which align with our model's bitrate settings.
    \item BigCodec \cite{xin2024bigcodec}: The official checkpoint\footnote{\url{https://github.com/Aria-K-Alethia/BigCodec}} with a bitrate of 1.04k bps was used.
    \item WavTokenizer \cite{ji2024wavtokenizer}: Two official checkpoints\footnote{\url{https://github.com/jishengpeng/WavTokenizer}}, WavTokenizer-large-600-24k-4096 (0.52k bps) and WavTokenizer-large-320-24k-4096 (0.975k bps) were used.
\end{itemize}

\subsubsection{Reproduced models based on official implementation}
Our primary baseline is BigCodec, as it is the current state-of-the-art single-codebook low-bps codec. 
We reproduced BigCodec based on its official implementation and further developed a streaming version (denoted as BigCodec-S)
by replacing all the non-causal convolution operations with causal convolution operations, and removing the upsampling/downsampling operations at the snake activation function\footnote{Removing the upsampling/downsampling operations yielided slighlty better results compared to using causal upsampling/downsampling operations in the streaming setting.}.
BigCodec-S was 
trained on the same datasets as TS3-Codec for a fair comparison.
We conducted extensive hyperparameter tuning (e.g., codebook size, loss weights) on BigCodec-S.
As a result,
at a similar bitrate, BigCodec-S trained using only Libri-light outperformed 
Mimi, the current state-of-the-art RVQ-based streaming codec model, on the majority of evaluation metrics, making our baseline strong.

\subsection{TS3-Codec model configuration}
We have trained several TS3-Codec models, (X1) - (X5), to fulfill different bitrates, as shown in Table~\ref{tab:tes_config}.
For TS3-Codec (X1), (X3) and (X5), the codebook size is 65536, resulting in bitrates of 640 bps and 800 bps.
For TS3-Codec (X2) and (X4), the codebook size is 131,072, resulting in lager bitrates of 680 bps and 850 bps.

\subsection{Training configurations}
All models were trained on 16 NVIDIA V100 32G GPUs. For BigCodec-S, the utterance length was set to 2.5 seconds, randomly cropped from the original utterance, and the batch size was adjusted to maximize GPU memory usage for different sizes of BigCodec-S models. Similarly, for TS3-Codec, the utterance length was set to 10 seconds, randomly cropped from the original utterance, and the batch size was adjusted to optimize GPU memory for different TS3-Codec models. We employed AdamW \cite{loshchilov2017decoupled} as the optimizer, with the moving average coefficients $\beta_1$ and $\beta_2$ set to 0.8 and 0.9, respectively. A scheduled learning rate was used, linearly declining from 1e-4 to 1e-5 for BigCodec-S and from 2e-4 to 2e-5 for TS3-Codec. We performed 1k learning rate warmup steps, and all models were trained for 500k steps.

\section{Experimental results}
\label{result}

\subsection{Results at approximately 1000 bps}

In Table~\ref{tab:1000bps}, four non-streaming baselines, (A)–(D1), are evaluated using results derived from their official checkpoints.
We have the following observations:
(1) Among the four non-streaming baselines, BigCodec (C) demonstrates the best performance at approximately 1000 bps, surpassing other codec models by a significant margin across all metrics. 
This superior performance is the reason we selected BigCodec as the basis for developing a streaming version as our baseline.
(2) WavTokenizer (D1) achieves good UTMOS scores, as highlighted in their paper, where they emphasize that reconstructed utterances from their models have strong naturalness. However, WavTokenizer (D1) exhibits poor WER performance, and WER is not reported in their paper.

Next, by comparing the six streaming models, we observe the followings:
(1) TS3-Codec models (X1 and X2 with sliding window size of 32) perform the best for STOI, PESQ, MCD, UTMOS, and the second best for WER and SPK-SIM.
(2). Mimi performs the best for WER, probably because of their inclusion of semantic distillation. We listened to utterances generated by Mimi, the audio quality is worse than the TS3-Codec.

Figure~\ref{fig:1000bps} compare TS3-Codec and BigCodec under the streaming settings, at a bitrate of approximately 1000 bps.
Note that at the 1000 bps bitrate setting, TS3-Codec operates at 800 or 850 bps, significantly lower than its counterpart, BigCodec-S, which operates at 1040 bps.
Comparing TS3-Codec and BigCodec-S, we can observe that:
(1) 
Under similar computational budgets,
the proposed TS3-Codec always outperforms BigCodec-S significantly across all metrics (e.g. in 640bps, UTMOS scores for TS3-Codec and BigCodec-S are 3.85 and 3.41, WERs are 3.6 and 4.6).
(2) 
TS3-Codec achieves comparable or superior performance 
to BigCodec with significantly less computation.
For example,
TS3-Codec, with just 12\% of the computation and 77\% of the bitrate of its BigCodec counterparts, similarly delivers comparable or superior results.

\subsection{Results at approximately 600 bps}

Table~\ref{tab:600bps} presents the results for bitrates around 600 bps. We observe the following:
(1). Under the streaming setup, TS3-Codec models (X3 and X4 with sliding window size of 16) achieve the best performance in STOI, PESQ, MCD, SPK-SIM, and UTMOS, while securing the second-best WER. TS3-Codec also outperforms the two non-causal baselines across all metrics.
(2). Mimi achieves the best WER, likely due to its incorporation of semantic distillation during training.
(3). SpeechTokenizer performs poorly in PESQ, MCD, SPK-SIM, and UTMOS metrics, though its WER is relatively decent. Upon listening, some male voices are distorted to sound like female robotic speech, yet the content remains intelligible. This may be due to SpeechTokenizer’s use of semantic distillation during training, allowing the first-stream tokens to encode sufficient information for content recognition.

At the 640 bps setting, TS3-Codec achieves comparable or superior performance to BigCodec while using only 15.6\% of the computation as shown in Figure~\ref{fig:640bps}. 

\section{Conclusion}
\label{Conclusion}
This paper introduces TS3-Codec, the first transformer-only NAC model designed for streaming processing with low computational complexity and a single codebook. 
Compared to BigCodec, the state-of-the-art convolutional codec, TS3-Codec achieved similar or better performance with just 12\% of the computation and 77\% of the bitrate, and performed significantly better under similar computational settings.
TS3-Codec sets a new direction for simple and efficient streaming NACs.

\bibliographystyle{IEEEtran}
\bibliography{mybib}

\end{document}